\documentclass[multi]{cambridge6A}
\usepackage[numbers]{natbib}
\usepackage{rotating}
\usepackage{floatpag}
\rotfloatpagestyle{empty}
\usepackage{amsthm}
\usepackage{graphicx,bm}

\usepackage{ubec}

\begin{document}

\author{R.A. Duine}
\affiliation{Institute for Theoretical Physics and Center for Extreme Matter and Emergent Phenomena, Leuvenlaan 4, 3584 CE, Utrecht, The Netherlands}

\author{Arne Brataas}
\affiliation{Department of Physics, Norwegian University of Science and Technology, NO-7491 Trondheim, Norway}

\author{Scott A. Bender and Yaroslav Tserkovnyak}
\affiliation{Department of Physics and Astronomy, University of California, Los Angeles, California 90095, USA}

\chapter{Spintronics and Magnon Bose-Einstein Condensation}

\begin{verbatim}
\end{verbatim}

\begin{abstract}
Spintronics is the science and technology of electric control over spin currents in solid-state-based devices. Recent advances have demonstrated a coupling between electronic spin currents in non-magnetic metals and magnons in magnetic insulators. The coupling is due to spin transfer and spin pumping at interfaces between the normal metals and magnetic insulators. In this Chapter, we review these developments and the prospects they raise for electric control of quasi-equilibrium magnon Bose-Einstein condensates and  spin superfluidity. 
\end{abstract}

\section{Introduction}
\label{sec:introduction}
In undergraduate texts, Bose-Einstein condensation is a phase transition that occurs in an ideal Bose gas at large enough densities. In developing set-ups to observe Bose-Einstein condensation in a way analogous to this textbook treatment, one faces several challenges. First of all, conserved bosons in condensed-matter systems are composite particles. For example, Cooper pairs of electrons are bosons but the Bardeen-Cooper-Schieffer regime of condensation (in which the electrons condense as a result of a weak and attractive effective interaction) is rather different from the physics of non-interacting point-like bosons \cite{PhysRev.106.162}. The size of the Cooper pairs, determined in the weakly-interacting limit by the coherence length $\xi \sim \epsilon_F/k_F \Delta$ with $k_F$ the Fermi wave number, $\epsilon_F$ the Fermi energy, and $\Delta$  the superconducting gap, is much larger than the distance between the pairs. This latter distance is estimated from the fraction $\Delta/\epsilon_F$ of the electrons that form Cooper pairs, so that the pair density is $n_p \sim k_F^3 \Delta/\epsilon_F \sim k_F^2/\xi $, and we have that $n_p \xi^3 \sim (k_F \xi)^2 \gg 1$.   

Cold atoms \cite{pethick_smith_book_02}, on the other hand, are composite bosons where the internal degrees of freedom are typically at much higher energies than their temperature, so that they can be considered as point particles. The bosonic atoms in magnetically-trapped ultracold atomic vapors are, however, not conserved and have a finite lifetime because atoms may escape from the trap (both as single atoms or after molecule-forming collision processes). The same mechanism also prevents the system from reaching its true thermodynamic ground state (a state that is most likely a solid).  Relaxing the requirement of strict boson conservation,  one may consider Bose-Einstein condensation as a quasi-equilibrium phenomenon. Prime examples of quasi-equilibrium Bose-Einstein condensation of non-conserved particles are condensates of photons \cite{klaers2010} and (exciton-)polaritons \cite{balili2007,kasprzak2010}. In true equilibrium, these particles would not be described by a Bose-Einstein distribution function with nonzero chemical potential and for both photons and polaritons the system is maintained in quasi-equilibrium at nonzero chemical potential by external pumping. The mechanism of reaching quasi-equilibrium is very different for each of these systems.

\begin{figure}
\includegraphics[width=10cm]{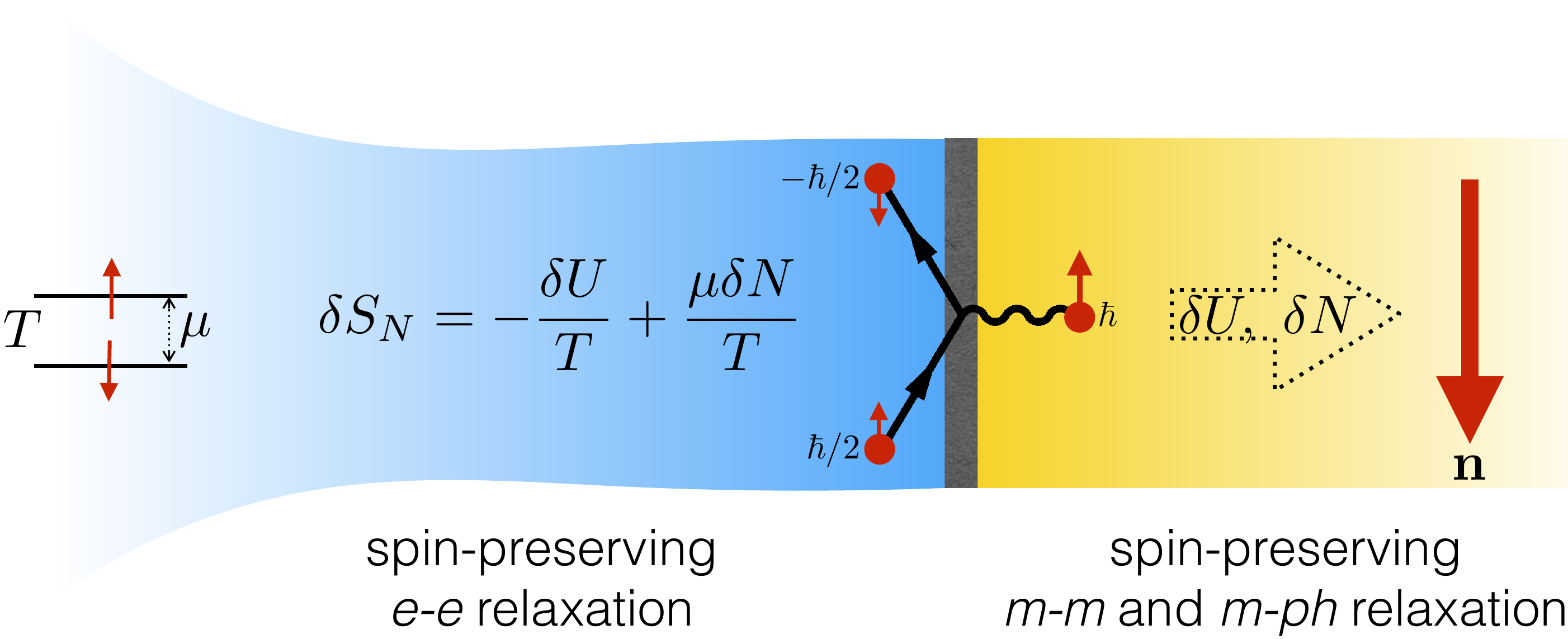} 
\caption{Schematic of a normal-metal/magnetic-insulator heterostructure, in which a finite chemical potential of magnons can be induced by a nonequilibrium electron-spin accumulation $\mu$. In the idealized case with no magnon-number relaxation in the insulator, the metallic reservoir acts as a bath supplying magnons and energy into the ferromagnet, with temperature $T$ and chemical potential $\mu$. $\delta S_N$ is the entropic change in the metal bath associated with the transfer of energy $\delta U$ and creation of $\delta N$ magnons in the ferromagnet (relative to the ordered spin orientation $\mathbf{n}$).}
\label{duine:fig:sch}
\end{figure}

In this Chapter we focus on magnons, quasi-particles corresponding to the quantized fluctuations of the magnetic order parameter, in magnetic insulators. These can undergo Bose-Einstein condensation as an equilibrium \cite{nikuniPRL00,ruegg2003} or a quasi-equilibrium \cite{demokritov2006} phenomenon. Here, we focus on quasi-equilibrium condensation of magnons. This is the result of excited magnons coming to quasi-equilibrium by magnon-conserving scattering processes, before they dissipate energy to the lattice and relax.  In experiments \cite{demokritov2006, serga2014} on quasi-equilibrium magnon condensation in solid-state magnetic insulators the excitation of magnons is achieved by microwave pumping. Recent advances in spintronics, however, have demonstrated interactions between magnons and electrons at interfaces between metals and magnetic insulators \cite{PhysRevLett.111.176601}. As we will discuss in detail in this Chapter, this opens the possibility of direct-current (DC) pumping of the magnetic system (see Fig.~\ref{duine:fig:sch}), and to achieve quasi-equilibrium magnon Bose-Einstein condensation in a solid-state DC transport experiment. On top of this, the integration of quasi-equilibrium magnon Bose-Einstein condensation with electronics opens the possibility for  electronic transport probes of the associated spin superfluidity.

In the remainder of this Chapter we first discuss the recent developments in spintronics that open the possibility of manipulating the magnetization in magnetic insulators by means of electrical current. Hereafter, we discuss the phase diagram of the DC-pumped quasi-equilibrium magnon gas. Finally, we briefly discuss signatures of the resulting spin superfluidity and give future prospects. 

\begin{figure}
\includegraphics[width=13cm]{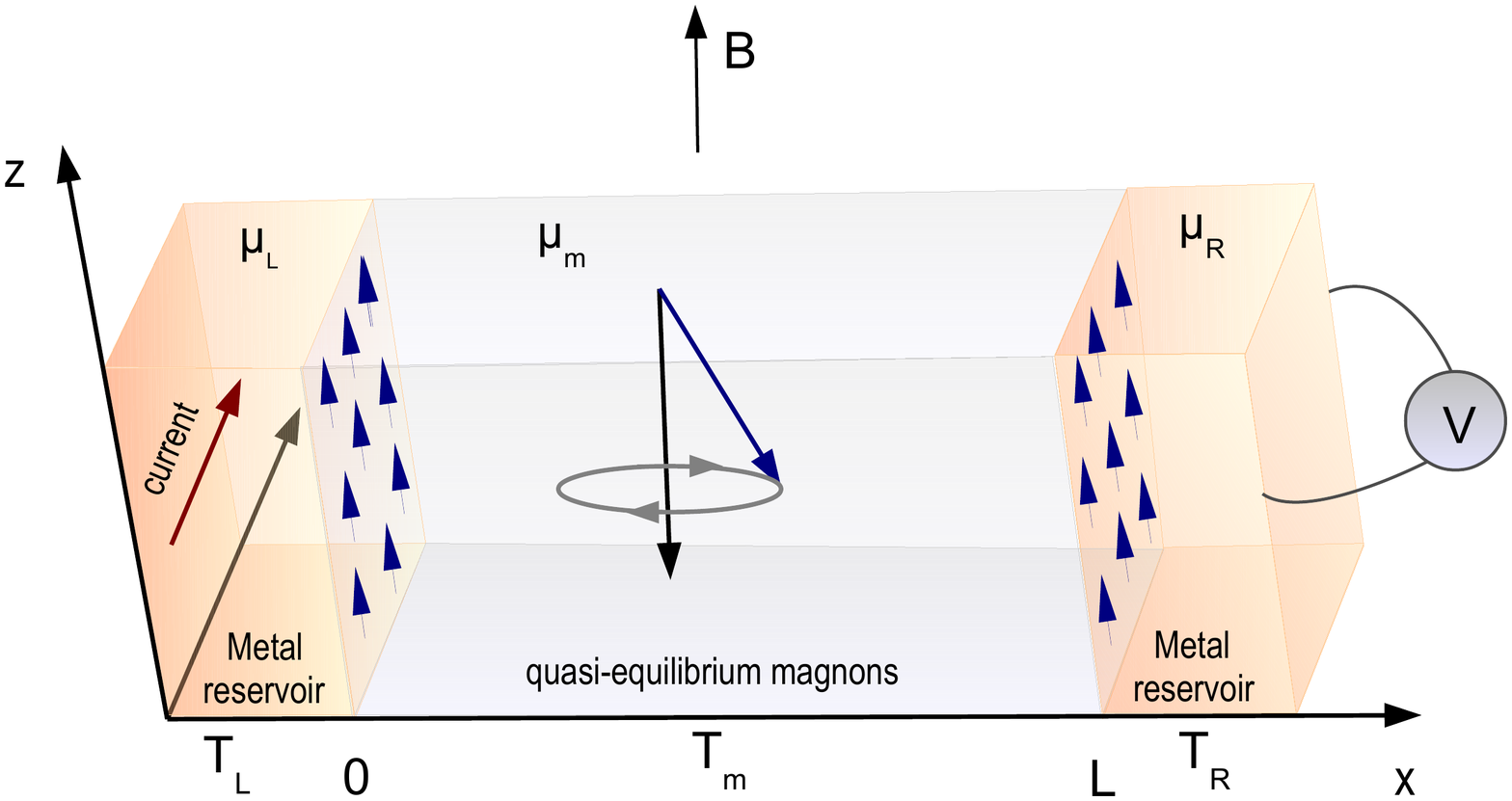} 
\caption{A set-up that we consider in this Chapter. A quasi-equilibrated magnetic insulator at temperature $T_m$ and chemical potential $\mu_m$ is connected to left and right by metallic reservoirs with spin accumulations $\mu_L$ and $\mu_R$, and temperatures $T_L$ and $T_R$, respectively. The spin accumulation in the reservoirs results from the spin Hall effect in response to currents in the $y$-direction (shown for left reservoir) and/or injected spin current from the magnetic insulator. Any spin current injected from the insulator to the reservoirs results in an inverse spin Hall voltage which can be measured (shown for right reservoir). The spin quantization axis is defined by the external field that we choose in the $z$-direction so that the equilibrium spin density points in the $-z$-direction.}
\label{duine:fig:device}
\end{figure}

\section{Spintronics}
\label{duine:sec:spintronics}
In this section we outline recent discoveries in spintronics that enable the integration of quasi-equilibrium magnon Bose-Einstein condensation with the control of spin currents. The first of these phenomena is the existence of spin currents across an interface between a normal metal and a magnetic insulator. The second concerns the generation of spin currents from charge current (or vice versa) via the (inverse) spin Hall effect. 

\subsection{Interfacial spin currents}
\label{duine:subsecinterfacespincurrent}

We consider the set-up in Fig.~\ref{duine:fig:device}, a magnetic insulator sandwiched between two normal metals. This geometry is chosen for convenience as it makes the theoretical analysis of the spin transport effectively one-dimensional. In an experiment it is easiest to put the metals on top of a thin film of a magnetic insulator. In terms of materials, an often studied system is the magnetic insulator Ytrrium-Iron-Garnett (YIG) interfaced with the non-magnetic normal metal platinum. We assume that the electrons in the normal metal on the left (right) have an electron spin accumulation that is nonzero at the normal-metal side of the interface between normal metal and magnetic insulator and is denoted by $\mu_L$ ($\mu_R$). This spin accumulation is an out-of-equilibrium electrochemical potential imbalance between up and down electrons, $\mu_L=\mu_\uparrow-\mu_\downarrow$ (with a similar definition of $\mu_R$). While in a more general set-up, the spin accumulation would be a vector, here we assume that only its $z$-component is finite, as illustrated by the electron spins in Fig.~\ref{duine:fig:device} in the normal metal. How the spin accumulation is established is discussed below. We further assume that magnons in the magnetic insulator are in quasi-equilibrium characterized by a magnon chemical potential $\mu_m$ and temperature $T_m$. The metallic leads are thermally biased, with $T_L$ and $T_R$ for the left and right reservoirs, respectively. 

We first focus on the left interface. At the interface, there exists an interface exchange coupling between the localized spins in the magnetic insulator, and the itinerant spins in the metal. Assuming this coupling is isotropic, a phenomenological expression is
\[
\hat V_{int}= \int d {\bf x} d {\bf x}' V ({\bf x},{\bf x}') \hat {\bf S} ({\bf x}) \cdot \hat {\bf s}( {\bf x}')~,
\]
where $\hat {\bf S} ({\bf x})$ corresponds to the spin density of the localized spins in the insulator, and $\hat {\bf s} ({\bf x})$ the spin density of the electrons in the metal. The latter is given by
\[
\hat   {\bf s} ({\bf x}) = \frac{\hbar}{2} \sum_{\sigma,\sigma' \in \{ \uparrow, \downarrow\}} \psi_\sigma^\dagger ({\bf x}) \bm{\tau}_{\sigma\sigma'} \psi_{\sigma'} ({\bf x})
\]
in terms of the Pauli matrices $\bm{\tau}$ and electron creation and annihilation operators $\hat \psi_\sigma^\dagger ({\bf x})$ and $\hat \psi_\sigma ({\bf x})$. Finally, the matrix elements $V({\bf x}, {\bf x}')$ decay rapidly away from the interface. 

We first assume that the spin density in the magnetic insulator can be treated classically and is homogeneous, and, moreover, that the temperature is relatively low so that we can approximate $\langle {\bf S} \rangle \simeq \hbar s {\bf n}$, with the density  $s=S/v$ where $S$ is the total spin per unit cell, $v$ the volume of a unit cell in the magnetic insulator, and ${\bf n}$ a unit vector in the direction of spin density. The $z$-component of the spin current from the magnetic insulator into the left reservoir, per unit area, is then given by \cite{PhysRevB.66.224403} (we define positive spin current as flowing to the right)
\begin{equation}
\label{duine:eq:spinpumping}
   j^{int}_s= -\frac{\hbar g_{\uparrow\downarrow}}{4 \pi} \left. {\bf n} \times \frac{d{\bf n}}{dt} \right|_z~,
\end{equation}
with $g_{\uparrow\downarrow}$ the so-called spin-mixing conductance (in units of m$^{-2}$ and disregarding its imaginary component).  As we shall see, this latter conductance characterizes the efficiency of spin transport across the interface. It can be straightforwardly calculated, e.g., by using perturbation theory in the coupling $V({\bf x},{\bf x}')$.  A microscopic expression is not needed at this point. The mixing conductance can also be determined from ab initio calculations \cite{jia2011} or from experiments \cite{PhysRevLett.111.176601}. For interfaces between YIG and Pt the mixing conductance is estimated to be up to $5$ nm$^{-2}$ depending on interface quality \cite{jia2011,burrowesAPL12}. We note that the above expression captures the spin-pumping contribution and not the spin-transfer contribution that results from the spin accumulation in the normal metal. 

We now consider the above expression for magnons. Using a linearized Holstein-Primakoff transformation, we have for the (circular) magnon annihilation operator that 
\[
   \hat b = \sqrt{\frac{s}{2}} \left(   \delta \hat n_x - i  \delta \hat n_y \right)~,
\]
where we assumed, as in Fig.~\ref{duine:fig:device}, the magnetic order to be in the $-z$-direction, so that ${\bf n} \simeq  (\delta \hat n_x, \delta \hat n_y,-1)$. Inserting this in the expression for the spin current leads to the replacement
\[
   \left. {\bf n} \times \frac{d {\bf n}}{dt} \right|_z\to \frac{4}{s} \frac{1}{V} \sum_{{\bf k}} n_{\bf k} \omega_{\bf k}~,
\]
with $\hbar \omega_{\bf k}$ the magnon dispersion, and where we inserted an additional factor of two to incorporate constructive interference of magnon modes at the interface \cite{PhysRevB.90.094409}. 
Here, we have made a Fourier transform so that the number of magnons at momentum ${\bf k}$ is $n_{\bf k} = \langle \hat b^\dagger_{\bf k} \hat b_{\bf k}\rangle $, and $V$ is the volume of the magnetic insulator. Furthermore, we normal-ordered the magnon creation and annihilation operators and have kept only the expectation values of  $\hat b^\dagger \hat b$ as we are interested in thermal magnons. Inserting this result in Eq.~(\ref{duine:eq:spinpumping}) yields in first instance
\[
      j^{ int}_s = -\frac{g_{\uparrow\downarrow}}{\pi s} \frac{1}{V}  \sum_{{\bf k}} n_{\bf k} \hbar \omega_{\bf k}~.
\]
We now consider the situation of quasi-equilbrium magnons at chemical potential $\mu_m$ and temperature $T_m$ so that 
\[
   n_{\bf k}  = n_B \left(\frac{\hbar \omega_{\bf k} - \mu_m}{k_B T_m}\right)~,
\]
with $n_B (x) = [e^x-1]^{-1}$ the Bose-Einstein distribution function. In equilibrium there is no spin current accross the interface. To account for this we generalize our treatment with the replacement (see Ref.~\cite{bender2012})
\[
    n_B \left(\frac{\hbar \omega_{\bf k} - \mu_m}{k_B T_m}\right) \to  n_B \left(\frac{\hbar \omega_{\bf k} - \mu_m}{k_B T_m}\right)  -  n_B \left(\frac{\hbar \omega_{\bf k} - \mu_L}{k_B T_L}\right)~.
\]
The first contribution corresponds to spin pumping, the emission of spin current from an excited magnet. The second term reflects spin transfer, the injection and absorption of spin current into a magnet. Over the last decade, both these phenomena have been investigated extensively for a classical magnetization, i.e., not including contributions of magnons (see e.g. Ref.~\cite{RevModPhys.77.1375} and Ref.~\cite{brataas2012}). 
Putting the above together, and rewriting momentum summations as energy integrations involving the magnon density of states $D(\epsilon)$, we finally find the result
\begin{equation}
\label{duine:eq:spincurrentmagnons}
j^{ int}_s = -\frac{g_{\uparrow\downarrow}}{\pi s} \int d\epsilon D (\epsilon) \left( \epsilon - \mu_L \right)
\left[ n_B \left(\frac{\epsilon - \mu_m}{k_B T_m}\right)  -  n_B \left(\frac{\epsilon - \mu_L}{k_B T_L}\right)\right]~.
\end{equation}
In linear response this yields 
\begin{equation}
\label{duine:eq:spincurrentlinearresponse}
  j^{int}_s =\frac{\sigma^{int}_s}{\hbar\Lambda} \left(\mu_L - \mu_m \right) + \frac{L^{int}_{SSE}}{\Lambda} \left( T_L - T_m \right)~,
\end{equation}
which defines an interface spin conductivity $\sigma_s^{int} = 3\zeta (3/2) \hbar g_{\uparrow\downarrow} /2\pi s\Lambda^2$ and the coefficient $L^{int}_{SSE} = 15\zeta (5/2) k_B g_{\uparrow\downarrow} /4\pi s\Lambda^2$.
Here, we assumed a gapless quadratic magnon dispersion $\hbar \omega_{\bf k} = J_s {\bf k}^2$ in terms of the spin stiffness $J_s$, so that the magnon density of states is $D(\epsilon) = \sqrt{\epsilon}/4\pi^2J^{3/2}_s$. Furthermore, we have introduced the length scale $\Lambda = \sqrt{4 \pi J_s/k_B T_m}$, the thermal DeBroglie wave length for the magnons. Eqs.~(\ref{duine:eq:spincurrentmagnons})~and~(\ref{duine:eq:spincurrentlinearresponse}) are the main expressions for the magnon spin current accross the interface that we will use below. The interface between the magnetic insulator and normal metal on the right can be treated analogously. The contribution proportional to the temperature difference in Eq.~(\ref{duine:eq:spincurrentlinearresponse}) and determined by the coefficient $L^{int}_{SSE}$ gives rise to an interface contribution to the so-called spin Seebeck effect, i.e., a spin current as a result of a temperature gradient \cite{uchida2010,PhysRevLett.106.186601}. This effect, and its reciprocal dubbed the spin Peltier effect \cite{PhysRevLett.113.027601}, are subject of intense investigation \cite{bauer2012}.

\subsection{spin Hall effect}
\label{duine:subsecspinHall}
Having addressed the spin current through the interface, we now turn to the question of what establishes the spin accumulations $\mu_L$ and $\mu_R$ in the normal metals on the left and right. These are the result of spin-current injection from the magnetic insulator into the normal metals, combined with spin current due to the spin Hall effect. The spin Hall effect arises due to spin-orbit coupling and is reflected as a spin current with a spin polarization and a spatial direction transverse to an applied electric field in a metal \cite{PhysRevLett.92.126603, murakami2003}. In the geometry of Fig.~\ref{duine:fig:device} the electric field (or charge current) is flowing in the $y$-direction, and gives rise to a spin current  that flows in the $x$-direction with spin polarization in the $z$-direction.
We now focus again on the left metallic reservoir and in first instance assume transport is diffusive with weak spin-orbit interactions (so that the scattering mean free path is shorter than the spin-flip diffusion length). Even though the diffusive approach is not generally applicable to experimental situations, as detailed below, it for now serves our pedagogical purpose. The set of equations that describes the coupled spin and charge dynamics in the normal metal are then given by (within our quasi-one-dimensional geometry) \cite{dyakonovJETP71}
\begin{eqnarray}
\label{duine:eq:spindrifteqs}
  j_c &=& \sigma E + \frac{\sigma_{SH}}{2 e} \partial_x \mu_L~, \nonumber  \\
 \frac{2e}{\hbar} j_s &=& -\frac{\sigma}{2e} \partial_x \mu_L - \sigma_{SH} E~.
\end{eqnarray}
In the above the electric field $E$ and charge current $j_c$ are in the $y$-direction and the electron charge is $e$. The charge conductivity $\sigma$ and spin Hall conductivity $\sigma_{SH}$ are both in units of $\Omega^{-1}$  m$^{-1}$. The second term in the first equation is the charge current that results from a gradient in the spin accumulation via the inverse spin Hall effect \cite{saitoh2006}. This is the Onsager reciprocal of the spin Hall effect and it is thus governed by the same coefficient $\sigma_{SH}$. The inverse spin Hall effect is a powerful means to detect spin current electrically, as discussed in more detail below.
Writing $\sigma_{SH}=\theta_{SH} \sigma$, the spin Hall effect is quantified in terms of the dimensionless quantity (dubbed spin Hall angle) $\theta_{SH}$. For Pt, $\theta_{SH}\sim0.05$ and for Ta its magnitude is similar, but the sign is opposite \cite{2011arXiv1111.3702L}. 

The above equations have to be complemented with a continuity-like equation,
\[
   \frac{\partial j_s}{\partial x} = - \Gamma \mu_L~,
\]
where the rate per unit volume $\Gamma$ phenomenologically expresses spin-flip relaxation in the metal on the left of the insulator. Insertion of the expression for the spin current into the latter equation yields the spin-diffusion equation
\begin{equation}
\label{duine:eq:diffusionmetal}
   \frac{\partial^2 \mu_L}{\partial x^2} = \frac{\mu_L}{\ell^2}~,
\end{equation}
with the spin-flip diffusion length of the left lead $\ell = \sqrt{\sigma \hbar/4e^2\Gamma}$. 

We now assume the interface between the left metallic reservoir and the magnetic insulator is at position $x=0$ and that the thickness of the metal is $L_x$ in the $x$-direction and $L_z$ in the $z$-direction. The equations for the spin current and the spin accumulation are solved with the boundary conditions $j_s (x=-L_x) =0$ and $j_s (x=0) = j_s^{int}$. Using the linear-response expression in Eq.~(\ref{duine:eq:spincurrentlinearresponse}) we find that the spin accumulation at the interface in the left reservoir is given by (assuming $\theta_{SH} \ll 1)$
\[
   \mu_L = \frac{\theta_{SH} I_c \hbar^2 \Lambda\left[1-\cosh \left ( \frac{L_x}{\ell}\right) \right]
+2eLL_z \left[ \sigma_s^{int} \mu_m +\hbar L_{SSE}^{int} (T_m-T_e)\right] \cosh \left ( \frac{L_x}{\ell}\right)
}
{2e L_x L_z \sigma_s^{int} \left[ \cosh \left( \frac{L_x}{\ell}\right)
+ \left(\frac{\hbar}{2e} \right)^2 \frac{\Lambda}{\ell} \frac{\sigma}{ \sigma_s^{int}} \sinh \left( \frac{L_x}{\ell}\right) \right]}~,
\]
where $I_c = j_c L_x L_z$ is the total current through the normal metal (where the current density is assumed to be homogeneous). Note that the above result shows that in the limit of no spin relaxation, i.e., $L_x \to\infty$, no net spin current is flowing across the interface as $j_s$  is constant and zero (according to the boundary condition at $-L_x$). Again we mention for completeness that an analogous derivation holds for the right metallic reservoir. 

\begin{figure}
\includegraphics[width=14cm]{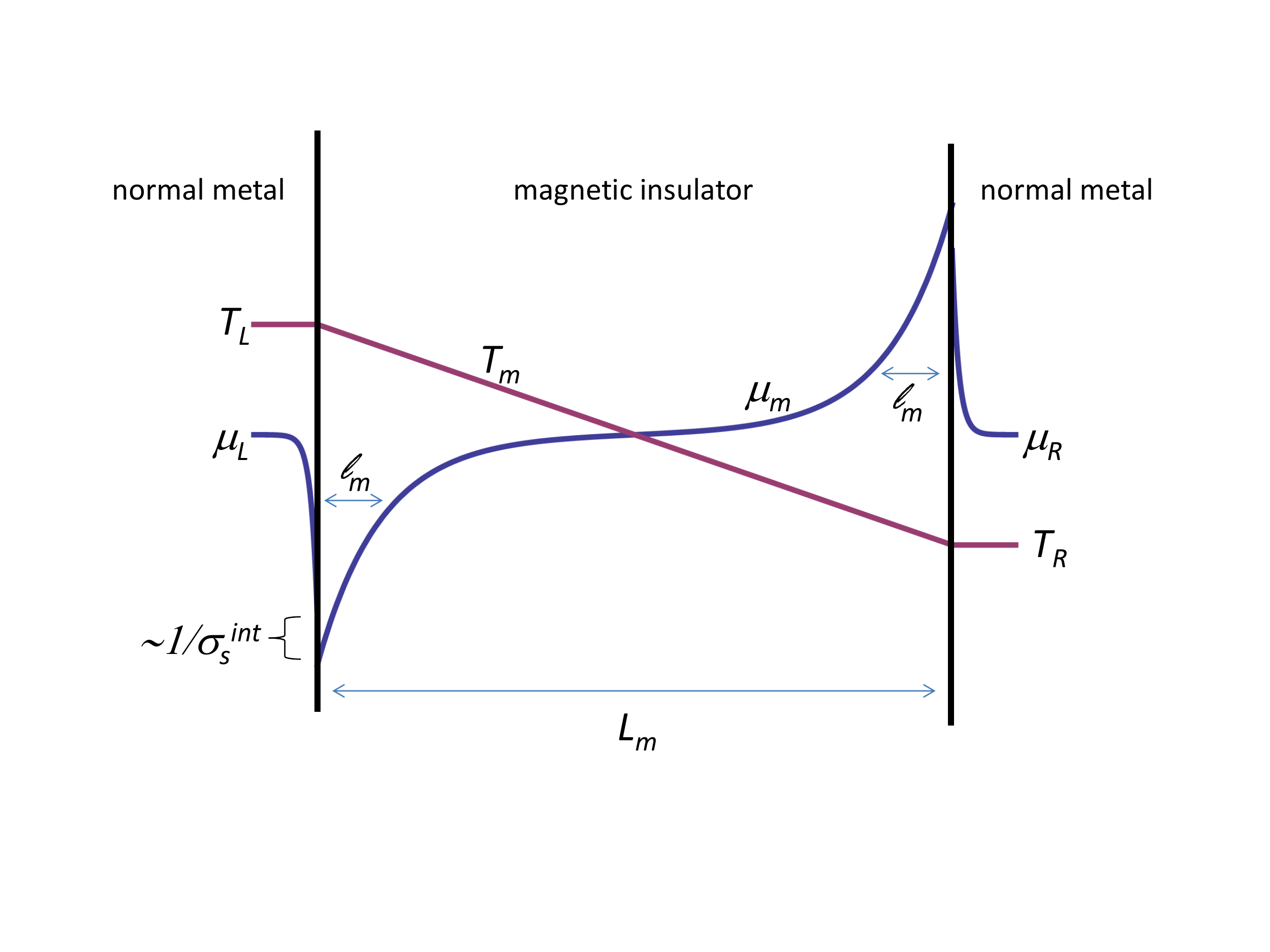} 
\caption{Spin accumulations $\mu_L$ and $\mu_R$, and magnon chemical potential $\mu_m$, that build up in the metallic reservoirs and magnetic insulator in response to a thermal gradient across the magnetic insulator. The spin accumulations and magnon chemical potential are nonzero over distances $\sim \ell$ and $\sim \ell_m$ (in metals and insulator, respectively) away from the interfaces. The jump across the interfaces is determined by the interface spin conductivity $\sigma_s^{int}$.}
\label{duine:fig:diffusion}
\end{figure}

\subsection{Inverse spin Hall voltage and the spin Seebeck effect}
\label{duine:subsec:spinseebeckeffect}
Spin current can be detected electrically via the inverse spin Hall effect as spin current injected into the metallic reservoirs gives a inverse spin Hall voltage in the $y$-direction across the reservoirs (in an open circuit geometry). As a further illustration, we extend the drift-diffusion treatment to capture the spin Seebeck effect. That is, we consider the inverse spin Hall voltage generated across the reservoirs as a result of a temperature gradient across the magnetic insulator. We assume, for simplicity, the temperature drops linearly $T_m (x) = (T_R-T_L)x/L_m + T_L$, with $T_L>T_R$ (see Fig.~\ref{duine:fig:diffusion}). We do not consider a temperature gradient across the metallic reservoirs, nor do we consider the effect of the temperature difference across the interface (governed by the Kapitza resistance).
In the diffusive limit of magnon transport, the magnetic insulator is described by similar equations as the diffusive normal metal, i.e., 
\begin{eqnarray}
  && j_s = - \frac{\sigma_s}{\hbar}\frac{\partial \mu_m}{\partial x} - L_{SSE} \frac{\partial T_m}{\partial x}~, \nonumber \\
  && \frac{\partial^2 \mu_m}{\partial x^2} = \frac{\mu_m}{\ell_m^2}~,
\end{eqnarray} 
where $\sigma_s$ is the spin conductivity of the magnetic insulator, $L_{SSE}$ its bulk spin Seebeck coefficient, and $\ell_m$ the thermal-magnon propagation length. Contrary to the case of Pt normal metals, a diffusive approach is expected to be appropriate for thermal magnons, albeit that the various microscopic details (such as the role of complicated spin-wave dispersions, strong magnon-phonon and magnon-magnon interactions) are not well-understood and may lead to a  non-trivial dependence of transport coefficients on magnetic field and temperature. The transport coefficients $L_{SSE}$ and $\sigma_s$ and the length scale $\ell_m$ are therefore neither experimentally nor theoretically very well understood for YIG at present. This, together with the complication of determining magnon temperature profiles from experimentally applied temperature differences \cite{PhysRevB.81.214418,PhysRevB.88.094410}, hinders a full and quantitative understanding of the spin Seebeck effect in YIG. Here our purpose is to discuss a simple example of diffusive spin transport that will be contrasted with the case of superfluid magnons later on.

We solve the spin diffusion equations in the magnetic insulator subject to the boundary conditions that the spin current vanishes at the left and right boundary of the system.  For both interfaces we use the boundary condition in Eq.~(\ref{duine:eq:spincurrentlinearresponse}) and we take zero charge current in both reservoirs. The properties of the metallic reservoirs, as well as their interfaces with the magnetic insulator, are taken equal.  Fig.~\ref{duine:fig:diffusion} shows a schematic plot of the spin accumulation and magnon chemical potential that build up in the metallic reservoirs  and magnetic insulator as a result of the thermal gradient. Within a distance $\sim \ell_m$ from the interfaces the magnon chemical potential is nonzero. Similarly, the spin accumulation in the metallic reservoirs is nonzero within a distance $\sim \ell$. The jump from spin accumulation to magnon chemical potential across the interface is inversely proportional to the interface spin conductivity $ \sigma_s^{int}$. 

The spin Seebeck coefficient is defined as $S_{SSE}=V_{ISHE}/(T_L-T_R)$, with the inverse spin Hall voltage $V_{ISHE}$ found by computing the electric field via Eqs.~(\ref{duine:eq:spindrifteqs}) from the injected spin current from magnetic insulator to metallic reservoir, and averaging over the $x$-direction. We consider now the limit $L_m \gg \ell_m$ which allows us to focus on one interface, which we choose to be the right one. Within the drift-diffusion theory the spin Seebeck coefficient is ultimately found as 
\begin{equation}
S_{SSE} = -\frac{\theta_{SH} \hbar L_y\ell_mL_{SSE}}{2eL_x L_m \left\{ \sigma_s^{int} + \left( \frac{\hbar}{2e}\right)^2 \left[ \frac{\ell_m}{\ell}\sigma + \frac{\Lambda}{\ell} \frac{\sigma \sigma_s}{\sigma_s^{int}}\right] \right\}} ,
\end{equation}
where $L_y$ is the length of the normal metals in the $y$-direction and where we also took the limit $L_x \gg \ell$. Within our drift-diffusion theory the spin Seebeck coefficient  (times $L_m$) saturates as a function of $L_m/\ell_m$ to the value determined by the above result. This saturation results from the fact that only magnons within a length $\ell_m$ from the interface contribute to the spin Seebeck voltage \cite{PhysRevB.88.064408,2013arXiv1306.0784K}. 

In the above treatment the properties of the insulator are characterized by the phenomenological constants $\ell_m$, $L_{SSE}$ and $\sigma_s$. The discussion is made more quantitative by computing these in the relaxation-time approximation in which $\sigma_s \sim  J_s \tau/\Lambda^3$ and $L_{SSE} \sim J_s k_B  \tau/\hbar \Lambda^3$, where $\tau$ is the magnon transport mean free time. Furthermore, the magnon spin propagation length is then given by $\ell_m\sim  v_m  \sqrt{\tau \tau_{sr}}$, where $v_m =2 \sqrt{J_s k_B T}/\hbar$ is the magnon thermal velocity and  $\tau_{sr}$ is the magnon spin-relaxation time. Here, $\tau$  is the result of various magnon conserving and non-conserving relaxation mechanisms such as magnon-phonon scattering, magnon-magnon (Umklapp) scattering, and scattering of magnons with impurities. The spin-relaxation time $\tau_{sr}$ acquires contributions only from processes that do not conserve magnon number. As we mentioned before, such relaxation and scattering processes are at present not fully understood. We remark that, by assuming Gilbert damping as the only relxation mechanism (which would be appropriate in clean systems at low temperatures), however, both time scales are on the order of $ \hbar/\alpha k_B T$, with $\alpha$ the Gilbert damping constant (which for YIG can be as small as $10^{-4}$). 

Because the spin diffusion length of Pt (which is on the order of a few nm at room temperature \cite{2011arXiv1111.3702L}) is not large compared to the mean free path, and because in experiments the Pt layers may be rather thin and the interface may be disordered, a diffusive bulk treatment of the transport in the normal metal is not generally applicable and the spin accumulation in the normal metal is in those situations not well defined. The appropriate variables are then the torque on the ferromagnet, and the (electrical) current in the normal metal. The relations between current and torques are then found from symmetry considerations and microscopic calculations \cite{PhysRevB.89.064425,PhysRevB.88.085423,PhysRevB.90.014428}, and one finds that the general phenomenology remains the same. Moreover, the system is described phenomenologically in terms of an effective spin Hall angle and effective mixing conductance that characterize coupling between magnetic dynamics and current, and loss of angular momentum at the interface, respectively \cite{PhysRevB.90.014428}.  The drift-diffusion theory discussed here gives an expression for these parameters in terms of the bulk spin Hall angle $\theta_{SH}$, spin-mixing conductance $g_{\uparrow\downarrow}$, and other parameters, that is appropriate if the diffusive treatment applies \cite{PhysRevB.90.014428}.

In summary, in this section we have established that temperature differences and differences between magnon chemical potential and electron spin accumulation drive spin transport across interfaces between normal metals and magnetic insulators. Furthermore, we have demonstrated how a current through a conductor that is parallel to its interface with a magnetic insulator sets up a nonzero spin accumulation via the spin Hall effect, and how spin current injected from a magnetic insulator to normal metal through an interface can be detected via the inverse spin Hall effect in the normal metal. 

We emphasize that the above analysis shows that the metallic reservoirs effectively act as grand-canonical baths for the magnons in the magnetic insulator (see Fig.~\ref{duine:fig:sch}), and that in this way the system we consider is rather close to the textbook grand-canonical treatment of Bose-Einstein condensation. In the next section we discuss how by tuning the driving forces, i.e., electrical currents and temperature differences, a quasi-equilibrium Bose-Einstein condensate of magnons can be achieved and maintained. 

\section{Pumping of quasi-equilibrium magnon condensation by spin current} 
\label{duine:sec:pumping}
In the previous section we have reviewed recent developments in spintronics concerning the linearized interaction between electrons and thermal magnons at interfaces between magnetic insulators and normal metals. In this section we consider the more general situation of a magnetic insulator that is partially condensed, as a result of interactions with the metallic reservoirs. 

To make the discussion concrete, we consider a magnetic insulator that is in the bulk described by a Heisenberg-model hamiltonian,
\begin{equation}
\label{duine:eq:heisenberg}
  \hat H [\hat {\bf S}] = - \frac{J}{2\hbar^2}\sum_{<i,j>} \hat {\bf S}_i \cdot \hat {\bf S}_j +  \sum_i \left[ \frac{K}{2\hbar^2} \hat S_{z,i}^2 +\frac{B}{\hbar} \hat S_{i,z}\right]~,
\end{equation}
with $J$ the nearest-neighbor exchange energy, $K>0$ the easy-plane anisotropy constant, and $B>0$ the external field in units of energy. At zero temperature and without dissipation, the dynamics of the average spin $\langle \hat {\bf S}_i \rangle \simeq \hbar S {\bf n}_i$ is governed by the Landau-Lifshitz equation
\begin{equation}
\label{duine:eq:LLeq}
    \frac{\partial \langle {\bf S}_i \rangle }{\partial t} = -\frac{1}{\hbar} \langle {\bf S}_i \rangle \times \frac{\partial \hat H [\langle \hat{\bf S} \rangle]}{\partial \langle {\bf S}_i \rangle }~,
\end{equation}
which describes precessional dynamics around the effective field.

The presence of the Bose-Einstein condensate is signalled by a nonzero expectation value $\Psi = \langle \hat b \rangle $ of the long-wavelength annihilation operator. Employing the same linearized Holstein-Primakoff transformation as in the previous section, we have that the magnetization direction for the condensed phase is at zero temperature given by ${\bf n} \simeq (\sqrt{2/s} {\rm Re} \Psi,-\sqrt{2/s} {\rm Im} \Psi,n_0/s-1)$ where $\Psi = \sqrt{n_0} e^{-i \varphi}$ with $n_0$ the condensate spin density and $\varphi$ the azimuthal angle of the magnetization with the $x$-axis.  
In the homogeneous situation, the Landau-Lifshitz equation then results in
\begin{equation}
\label{duine:eq:josephson1}
  \hbar \frac{d \varphi}{dt} = \mu_c~,
\end{equation}
with the condensate chemical potential $\mu_c = B-KS+KSn_0/s$. The easy-plane anisotropy leads to a mean-field self-interaction for the condensate. Within the Landau-Lifshitz description, the condensate density is time independent. We now incorporate spin injection at the interface and magnetization relaxation to derive a rate equation for the condensate density. 

\subsection{Condensate rate equation}
\label{duine:sec:condensatedynamics}
For simplicity, we consider the situation as depicted in  Fig~\ref{duine:fig:device}, and focus only on the right reservoir. The interface spin current, determined by Eq.~(\ref{duine:eq:spinpumping}), is then given by 
\[
     \frac{\hbar g_{\uparrow\downarrow}n_0}{2 \pi s} \frac{d \varphi}{dt}~.
\]
 Using Eq.~(\ref{duine:eq:josephson1}), and substituting $\mu_c \to \mu_c-\mu_R$ to account for the nonzero spin accumulation in the right reservoir, this result is rewritten to yield the condensate contribution to the spin current across the interface between metal and insulator \cite{bender2012}, 
\begin{equation}
\label{duine:eq:bccondensate}
  j_{s,c}^{int} = \frac{g_{\uparrow\downarrow} n_0}{2 \pi s} \left( \mu_c -\mu_R \right)~.
\end{equation}
The condensate spin current is accompanied by the spin current carried by thermal magnons, which, using Eq.~(\ref{duine:eq:spincurrentmagnons}), is found to be
\begin{eqnarray}
\label{duine:eq:spincurrentmagnonsright}
&& j^{ int}_{s,x} (\mu_m,\mu_R,T_m,T_R) = \frac{g_{\uparrow\downarrow}}{\pi s} \int d\epsilon D (\epsilon) \left( \epsilon - \mu_R \right)
\nonumber \\ 
&& \mbox{   }\times \left[ n_B \left(\frac{\epsilon - \mu_m}{k_B T_m}\right)  -  n_B \left(\frac{\epsilon -  \mu_R}{k_B T_R}\right)\right]~.
\end{eqnarray}
In the above expressions for both condensate and thermal magnon spin currents, the first term corresponds to spin pumping while the second accounts for spin transfer. 

Relaxation processes are at small energies accurately described by the Landau-Lifshitz-Gilbert phenomenology \cite{1353448}. This implies that the equation for the dynamics of the magnetization direction in Eq.~(\ref{duine:eq:LLeq}) acquires a Gilbert damping term parametrized by a dimensionless Gilbert damping constant $\alpha$, given by
\[
   \left. \frac{d{\bf n}}{dt} \right|_{rel} = - \alpha {\bf n} \times \frac{d {\bf n}}{dt}~. 
\]
For the condensate, this results in a loss of condensed magnons according to 
\begin{equation}
\label{duine:eq:gilbertdampingcondensate}
   \frac{dn_0}{dt} =- \frac{2 \alpha n_0 \mu_c}{\hbar}~.
\end{equation}
This loss term is similar in form as the contribution from spin pumping, as spin pumping corresponds to loss of angular momentum from the magnetic insulator to the normal metal across the interface. 

When the magnons are partially condensed, the density of thermal magnons is fixed and any spin current entering the magnetic insulator through its interface with the right reservoir is eventually absorbed by the condensate \cite{bender2012}. We now assume that this absorption is instantaneous. Physically, this corresponds to the regime where the scattering rate due to interactions between the thermal cloud and condensate is fast compared to the rate of magnon absorption/excitation at the interface, and to the damping rate. We denote $j_x \equiv - j^{ int}_{s,x} (\mu_c,\mu_R,T_m,T_R)$ which is (minus) the interface spin current of thermal magnons in the partially condensed phase. The density of condensed magnons is then determined by the equation
\begin{equation}
\label{duine:eq:rateeqcondensate}
  \frac{dn_0}{dt} = \frac{j_x}{\hbar L_m} - \frac{g_{\uparrow\downarrow} n_0}{2 \pi s L_m}\frac{ \left( \mu_c -\mu_R \right)}{\hbar} - \frac{2 \alpha n_0 \mu_c}{\hbar} \equiv \frac{j_x}{\hbar L_m} - \frac{n_0}{\tau_c}~,
\end{equation}
having defined
\[
   \frac{1}{\tau_c} = \frac{1}{\tau_0} - \frac{g_{\uparrow\downarrow}KSn_0}{2\pi s^2 \hbar L_m } - \frac{2 \alpha KS n_0}{\hbar s}~, 
\]
where $\tau_0$ denotes $\tau_c$ in the limit $n_0 \to 0$. For temperatures $k_B T \gg K$ the thermal spin current $j_x$ is to a good approximation independent of $n_0$ \cite{PhysRevB.90.094409}.

\begin{figure}
	\includegraphics[width=9cm]{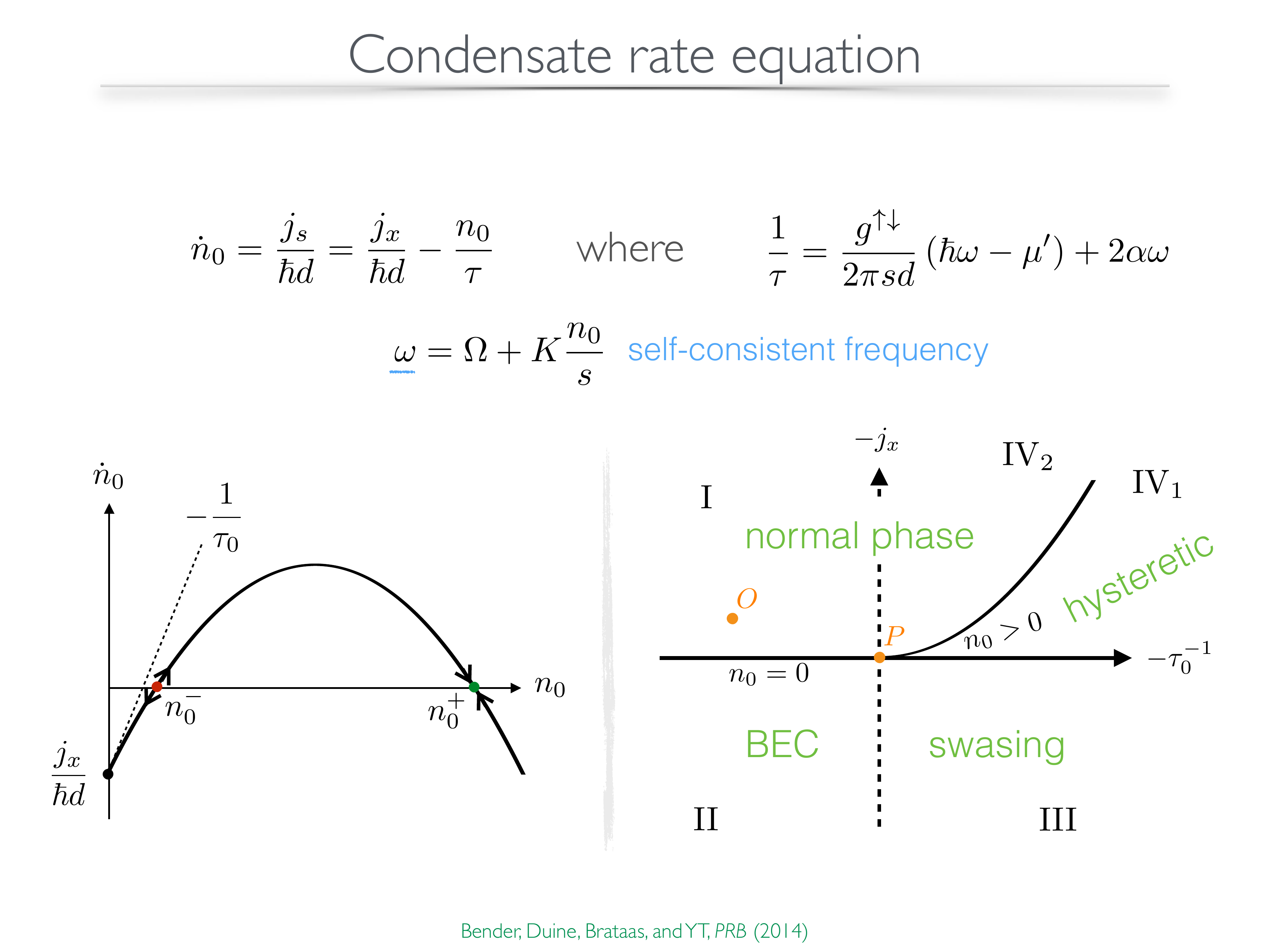} 
	\caption{Phase diagram of a quasi-equilibrium Bose-Einstein condensate maintained by spin-current injection. The point $O$ corresponds to thermal equilibrium, i.e., no driving forces. The possible phases that meet at the critical point $P$ are separared by transitions (solid lines) or cross-overs (dashed lines). For further details see main text. }
	\label{duine:fig:phasediag}
\end{figure}

The four possible choices of absolute and relative signs of $\tau_0$ and $j_x$ give rise to four distinct quadrants in the steady-state ($dn_0/dt=0$) phase diagram (see Fig.~\ref{duine:fig:phasediag}): In region I both condensate and thermal spin current lead to loss of magnons and prevent formation of the condensate. In region II the condensate spin current leads to decay of condensate magnons, which are replenished from the thermal cloud by injection of thermal magnons. This results in a steady state in which a condensate exists. Region II crosses over to region III which is the swasing regime \cite{PhysRevB.54.9353}, defined by the spin injected directly into the condensate as the spin accumulation is then larger than the magnon ground state energy. In this crossover the number of condensed magnons becomes larger, more rapidly so at lower temperatures. Finally, in region IV$_2$  the condensate spin current is not large enough to overcome the losses due to spin current by thermal magnons. In region IV$_1$, on the other hand, two steady-state solutions exist. Only one of these is stable, however, and hence the number of condensate magnons needs to be above a critical number in order for the condensate to be maintained. This leads to hysteretic behavior in the first-order transitions from regions IV$_2$ and III to region IV$_1$. The phase diagram in terms of the spin accumulation $\mu_R$ and the magnon and electron temperatures is determined by calculating $j_x$ and $\tau_0$ in terms of them, and reported in Ref.~\cite{PhysRevB.90.094409}. We conclude that quasi-equilibrium magnon Bose-Einstein condensation can be achieved by appropriate tuning of these driving forces. 

\subsection{Spin superfluidity} 
\label{duine:subsec:spinsuperfluidity}
One of the most striking consequences of Bose-Einstein condensation of magnons is the resulting spin superfluidity \cite{PhysRev.188.898,doi:10.1080/00018731003739943}. Developing a theory for spin transport across the pumped magnon system is not straightforward, as the phase diagram itself will change in an inhomogeneous situation with respect to the homogeneous case discussed so far. Here, we ignore such issues and discuss spin superfluidity for the case of an equilibrium condensate. For the model hamiltonian in Eq.~(\ref{duine:eq:heisenberg}), an equilibrium condensate forms at low temperatures when the external field is small enough so that the easy-plane anisotropy tilts the magnetization away from the $z$-axis. The general phenomenology of the superfluid spin transport is expected to be similar when comparing equilibrium and pumped condensates. 

The spin superfluid transport is for a small precession angle and at low temperatures conveniently described by the continuity equation for the density of condensed magnons (that includes a loss term due to Gilbert damping), and the Josephson equation for the condensate phase. These are found from the Landau-Lifschitz-Gilbert equation [Eq.~(\ref{duine:eq:LLeq}) with the Gilbert damping added] and in the long wavelength limit given by (at zero temperature)
\begin{eqnarray}
  \hbar \frac{dn_0}{dt} &=& - \nabla \cdot {\bf j}_s - 2 \alpha n_0 \mu_c~; \nonumber \\
  \frac{d {\bf v}_s}{dt} &=& - \frac{2 J_s \nabla \mu_c}{\hbar^2}~,
\end{eqnarray}
with ${\bf v}_s = -2 J_s \nabla \varphi/\hbar$ the superfluid velocity and ${\bf j}_s = n_0 \hbar {\bf v}_s$ the condensate (superfluid) spin current. In the above we assume the condensate to be spatially homogeneous, and that $n_0 \ll s$ and $\alpha \ll 1$. The stiffness $J_s=J S a^2/2$, with $a$ the lattice constant. For the configuration in Fig.~\ref{duine:fig:device} we find, by applying the boundary condition in Eq.~(\ref{duine:eq:bccondensate}) at both reservoirs, that the spin current injected into the right reservoir is given by 
\[
  j_s =\frac{n_0}{4\pi s} \left( \frac{g_{\uparrow\downarrow}^2\mu_L}{g_{\uparrow\downarrow}+2\pi \alpha s L_m} \right)~,
\]
where we took $\mu_R=0$.
Consequently, spin superfluidity is signalled by an algebraic decay of the spin current, and the ensuing inverse spin Hall voltage, as a function of the size of the magnetic insulator in the spin current direction \cite{doi:10.1080/00018731003739943,PhysRevLett.112.227201,PhysRevB.90.220401}. This is in contrast to the exponential decay (with the length scale $\ell_m$) that is found in the normal state \cite{PhysRevLett.109.096603}. The algebraic decay is a consequence of the small but nonzero Gilbert damping. In fact, the above shows that in the case when damping is absent ($\alpha=0$) the spin  current through a spin superfluid magnetic insulator is limited only by the interface (via the mixing conductance $g_{\uparrow\downarrow}$).

For the geometry of our example, spin superfluidity may be pinned by anisotropies that break the rotation symmetry around the $z$-axis \cite{doi:10.1080/00018731003739943,PhysRevB.90.220401} (which ultimately gives to the $U(1)$ symmetry that is spontaneously broken by the condensation). As an example, we consider the addition of a term $-K_x \sum_i\hat S_{i,x}^2/2 \hbar^2$ to the hamiltonian in Eq.~(\ref{duine:eq:heisenberg}). With this addition, the spin hamiltonian for constant density reduces to the energy 
\begin{equation}
  E = \int  d {\bf x} \left[ J_s n_0 \left( \nabla \theta\right)^2 + (B-KS)n_0 + \frac{KSn_0^2}{2s}- K_x S n_0 \cos^2 \theta \right]~,
\end{equation}
where we again took the long wavelength limit. The spin-current-carrying state requires gradients in the phase. These are penalized by the term $\sim K_x$ which pins the phase at $\theta=0$ or $\pi$. The competition between exchange and in-plane anisotropy in the above energy thus defines a lower critical current
\[
   j_{c,low} = 2 n_0 \sqrt{J_s S K_x}/\hbar~,
\]
below which spin superfluid flow is pinned. Physically, this lower critical current follows from comparing the energy of the current-carrying state to the energy of a domain wall in $\theta$ from, e.g., $0$ to $\pi$. Once the energy of the current-carrying state exceeds this domain wall energy, domain walls, and therefore gradients in the phase, are created, thus allowing finite superfluid spin currents. Below this lower critical current the anisotropy makes it energetically favorable for the phase to remain homogeneous away from the interface \cite{doi:10.1080/00018731003739943,konigPRL01}.

The upper critical current is found by realizing that the condensate density is actually $n_0=s(1-n_z^2)/2$ in terms of the $z$-component of the magnetization direction. From the above energy we then find that the superfluid flow is unstable towards increasing $n_z$ for $|\nabla \theta| \sim B-KS$, which defines an upper critical current. When the current approaches this critical value, the superfluid-carrying state is relaxed by spontaneous vortex-induced phase slips transverse to our quasi-one-dimensional geometry \cite{doi:10.1080/00018731003739943}.
 
\section{Perspectives}
\label{duine:sec:concl}
In this Chapter we have discussed how spin currents flowing across the interface between a magnetic insulator and normal metal can be used to achieve quasi-equilibrium Bose-Einstein condensation of magnons in the magnetic insulator, and, moreover, to probe its spin transport properties. Future theoretical works should improve on our treatment of interactions, in particular regarding the interactions between condensate and thermal magnons and the role of phonons. Moreover, the description of the condensed phase may require inclusion of the dipolar interactions. Another direction for study are inhomogeneous situations and in particular the coupled spin-heat transport properties of the partially-condensed magnon system. 

On the experimental side, the best-studied systems are devices with only one metallic reservoir and where the magnetic insulator is YIG and the metal is Pt. More complicated set-ups involving more than one reservoir have not yet been seriously addressed. Moreover, from a materials-science perspective there is ample room for exploring novel materials and optimizing interface spin transport properties, e.g., by considering antiferromagnets \cite{PhysRevB.90.094408}. In conclusion, we expect that the interplay between magnonic many-body physics and spintronics will be a source of new physics in the years to come. 

 This work was supported by the Stichting voor Fundamenteel
Onderzoek der Materie (FOM) and is part of the D-ITP consortium,
a program of the Netherlands Organization for
Scientific Research (NWO) that is funded by the Dutch
Ministry of Education, Culture and Science (OCW) (R.A.D.), by the European
Research Council (ERC) (R.A.D and A.B.), by the EU-FET grant InSpin 612759 (A.B.), and by the US DOE-BES under Award No. DE-SC0012190 (S.A.B. and Y.T.).

\bibliography{SpintronicsAndMagnonBEC}
\bibliographystyle{cambridgeauthordate}

\end{document}